%% file: acmmanuscript.tex
\PassOptionsToPackage{twoside=false}{geometry}
\documentclass[acmsmall,screen,nonacm]{acmart}

\usepackage{graphicx}
\usepackage{listings}
\usepackage{comment}
\usepackage{listofitems}
\usepackage{booktabs} 
\usepackage{enumitem}
\usepackage{multirow}
\usepackage{eucal}
\usepackage{array}
\usepackage{makecell}
\usepackage{hyperref}
\usepackage{diagbox}
\usepackage{threeparttable}
\usepackage{pifont}
\usepackage{tcolorbox}
\usepackage{siunitx}
\usepackage{balance}
\AtBeginDocument{%
  }

\begin{document}

\title{Understanding Python Linters through a Comprehensive and Systematic Empirical Study}



\author{Qingyun Yang}
\email{qingyun.yang@smail.nju.edu.cn}
\affiliation{%
  \institution{Nanjing University}
 \city{Nanjing}
  \state{Jiangsu}
  \country{China}
}

\author{Zejun Zhang}
\authornote{Corresponding author.}
\email{zejun.zhang@nju.edu.cn}
\affiliation{%
  \institution{Nanjing University}
  \city{Nanjing}
  \state{Jiangsu}
  \country{China}
}

\author{DAPENG YAN}
\email{dapeng.yan@njupt.edu.cn}
\affiliation{%
  \institution{Nanjing University of Posts and Telecommunications}
  \city{Nanjing}
  \state{Jiangsu}
  \country{China}
}

\author{Tian Zhang}
\email{ztluck@nju.edu.cn}
\affiliation{%
  \institution{Nanjing University}
  \city{Nanjing}
  \state{Jiangsu}
  \country{China}
}
\author{Zhenchang Xing}
\affiliation{%
  \institution{CSIRO's Data61}
  \city{Canberra}
  \country{Australia}}
\email{zhenchang.xing@data61.csiro.au}

\author{Yi Li}
\email{yi\_li@ntu.edu.sg}
\affiliation{%
  \institution{Nanyang Technological University}
  \city{Singapore}
  \country{Singapore}
}

\author{Zhiming Liu}
\email{zhimingliu88@swu.edu.cn}
\affiliation{%
  \institution{Southwest University}
  \city{Chongqing}
  \country{China}
}






\renewcommand{\shortauthors}{Yang et al.}

\begin{abstract}
  Code linters, tools that automatically detect code issues, help developers accelerate development and improve code quality by identifying errors early. 
Unlike previous studies that focused mainly on single linters or informal comparisons, this paper presents the first large-scale empirical study of five widely used Python linters: Ruff, Pylint, Flake8, Mypy and Black, through four key user concerns: linting rule checks, speed, configuration methods, and linter preferences. 
We analyze 1,551 linting rules, categorize them into five types, and construct rule mappings, finding that more than 40\% are shared between multiple linters, with Ruff and Pylint together covering over 80\% of all rules. 
Pylint has the most rules in the first two categories, and Ruff leads in the last three. 
Next, we measure their running times on 1,000 GitHub projects under three setups (single-rule, default, all-rules), finding a consistent speed ranking from fastest to slowest: Ruff, Mypy, Flake8, Black, and Pylint. 
Ruff is extremely fast, with runtimes typically within one second. 
Then, we summarize four configuration methods and four types of configuration options from 443 available options across the five linters. 
While all linters support the four configuration methods, they differ in project configuration formats and flexibility in their configuration options. 
Finally, an analysis of 3,094 GitHub projects using the five linters reveals that over 50\% use multiple linters, share strong consensus on option selections and rule suppressions, but exhibit project-specific variations in enabled-rule configurations. 
Our findings provide actionable insights for linter users and establish a benchmark dataset and taxonomy to support future research, such as cross-linter rule mapping, performance modeling, and automated configuration.
\end{abstract}





\maketitle

\section{Introduction}
\input{intro}

\section{Study Design}\label{motivation}

\input{motivation}


\section{EMPIRICAL ANALYSIS}\label{result}
 \input{result_four_rqs_no_features}
\section{Discussion}\label{discuss}
\input{discussion}

\section{Related Work}\label{relatedwork}
\input{related}
\section{Conclusion and Future Work}\label{conclusion} 
This paper presents both qualitative and quantitative analyses of linting rule checks, speed, configuration methods, and linter preferences across five Python linters: Ruff, Pylint, Flake8, Mypy, and Black. 
Our findings indicate that no single linter excels in all aspects, highlighting the need for Python linter users to choose based on their specific project requirements. 
In the future, we plan to expand our study to include more linters and automatically select and generate configurations based on users' requirements. 
The replication package~\cite{data_code} is available for further research. 
\bibliographystyle{ACM-Reference-Format}
\bibliography{sample-base}










\end{document}

%% file: intro.tex
Linters are automatic static analysis tools that detect code issues (e.g. stylistic issues and violations of coding standards)~\cite{tomasdottir2020adoption,relationship_linter_maintain,liter_use_warn}. 
They play an important role in software development by improving developer efficiency, enforcing consistent coding practices, and ensuring code quality~\cite{jaspan2007understanding,tomasdottir2017and}. 
To support different programming ecosystems, a variety of linters~\cite{black,eslint,ckstyle} are developed and widely used. 
For instance, Ruff and Pylint for Python, ESLint and JSHint for JavaScript. 

Previous studies~\cite{latappy2023mlinter,eghbali2025dylin,hart2023eastwood,oumarou2015identifying,tsantalis2008jdeodorant,loriot2022styler,vayadande2023let,holden2024code,walchshofer2024introducing,beller2016analyzing,zampetti2017open,christakis2016developers,tomasdottir2020adoption,tomasdottir2017and,liter_use_warn,relationship_linter_maintain} typically focus on how developers use a single linter but overlook a comparative analysis of linters about aspects where users face challenges, such as rule checks.  
With many linters available and little guidance on how to select and use linters, a systematic comparison is essential to help users make informed choices. 
However, such comparative study is labor-intensive, requiring deep familiarity with each tool and careful cross-linter comparison. 
For example, collecting, analyzing and comparing hundreds of linting rules across linters demands substantial manual effort. 
Tómasdóttir et al.~\cite{tomasdottir2020adoption} interviewed developers to understand the adoption of ESLint and found that users struggle to understand and choose rule checks. 
Their findings show the need for in-depth analysis of linting rules within a linter, let alone across multiple linters, further highlighting the necessity of such comparative study. 

To systematically conduct the comparative study of linters, we first select five representative Python linters with high popularity and comprehensive rule coverage: Ruff~\cite{ruff}, Pylint~\cite{pylint}, Flake8~\cite{flake8}, Mypy~\cite{mypy}, and Black~\cite{black}.
We focus on Python linters due to Python's top ranking in popularity in recent years~\cite{ranking_pl_ieee,popular_pl_tiobe}. 
Next, to understand the challenges users face when using linters, we examine discussions on Stack Overflow~\cite{stack_overflow}, which tend to focus on technical issues, and Reddit~\cite{reddit}, which often reflects broader user experiences. 
We find that users are frequently confused by four aspects of these linters: \textbf{rule checks, speed, configuration methods, and linter preferences}. 
For example, due to the variety of linting rules across linters, a Stack Overflow user mistakenly believed that Mypy and Pylint cannot check undefined object~\cite{so_checks}. 
In reality, Mypy excels at object usage and includes support for such checks. 
The result underscores the need to analyze and compare linters across these aspects, helping users correctly understand and utilize linters.

Motivated by these user concerns, our study about five Python linters addresses the following four research questions: 

\textbf{RQ1: What linting rules are enforced by Python linters?} We first collect linting rules from linter official documentation~\cite{ruff_rule,pylint_rule,flake8_rule,mypy_rule,black_rule} and then categorize them using a card sorting approach~\cite{card_sorting}.  
Among the 1,551 linting rules,  56\% are unique to a single linter and no linting rules exist in all linters, reflecting linters have their distinctiveness. 
These rules fall into five categories: \textit{wrong syntax}, \textit{wrong usage of objects}, \textit{logical contradiction}, \textit{style convention} and \textit{best practice}. 
Pylint has the most rules in the first two categories compared to other linters, and Ruff has the most rules in the last three. 
To maximize linting coverage with minimal linters, using only Ruff and Pylint can cover over 80\% of linting rules. 
The large number of linting rules and their overlaps emphasize the importance of building linter interoperability to enable linters to work seamlessly together, reduce redundant checks and improve efficiency. 

\textbf{RQ2: What is the running time of Python linters on projects?} To comprehensively assess the linter running time, we conduct three setups: enabling a single rule, default rules and all rules, corresponding to minimum, typical and maximum running time of linters, respectively. 
Based on rigorous runtime measurements on 1,000 GitHub projects under three setups, the linters consistently rank from fastest to slowest: Ruff, Mypy, Flake8, Black and Pylint, 
with runtimes of 5.1ms $\sim$ 1s, 55.9ms $\sim$ 2s, 50.3ms $\sim$ 5.5s, 61.3ms $\sim$ 33.7s, and 110.6ms $\sim$ $>$1000s, respectively. 
Except for Black and Flake8, the running times of different rules exhibit significant differences, indicating that linter running time is affected by rule selection.

\textbf{RQ3: How are Python linters configured?} By analyzing linter official configuration documentation~\cite{ruff_config,pylint_config,flake8_config,mypy_config,black_config}, we identify four common configuration methods: \textit{command line}, \textit{project configuration files}, \textit{dedicated configuration files}, and \textit{inline configuration}. 
All five linters support these methods, reflecting a shared emphasis on flexible integration into diverse development workflows. 
However, they differ in the formats of project configuration files. 
To further examine configuration flexibility, we collect 443 configuration options and categorize them into four types: \textit{environment setup}, \textit{scope control}, \textit{rule specification} and \textit{result handling}. 
Ruff, Pylint and Mypy each offer extensive category support, whereas Flake8 and Black offer only 25 options each, reflecting a more opinionated and minimalist design philosophy. 
Given the complexity of configuration methods and options across linters, users should consider the flexibility and level of control they need for their projects before selecting suitable linters. 


\textbf{RQ4: What are the preferences for selecting linters, configuring linters, and choosing linting rules across projects?} We collect 3,094 GitHub projects using the five Python linters. 
Over 50\% of projects use multiple linters, reflecting a preference for comprehensive rule checks. 
Most projects using Ruff, Mypy, Flake8 and Black,  prefer project configuration files, while projects using Pylint predominantly use dedicated configuration files. 
Projects' configuration option preferences focus primarily on \textit{rule specification}, followed by \textit{scope control}, whereas \textit{environment setup} and \textit{result handling} feature less prominently among commonly configured options.
For linting rules, Black restricts enabling and disabling rules, reflecting its opinionated design. 
While enabling rules, most linters (Pylint, Flake8, Mypy) reflect project-specific preferences; when disabling, most linters (Ruff, Pylint, Flake8) show a consensus, focusing on suppressing \textit{style convention} checks. 
This highlights the challenge of effectively using linters and emphasizes the importance of automatically recommending personalized rule selections and configurations based on project requirements. %

The contributions of this paper are as follows:
\vspace{-1mm}
\begin{itemize}[leftmargin=*]

    \item To the best of our knowledge, this is the first large-scale qualitative and quantitative study that analyzes and compares linting rule checks, speed, configuration methods, and linter preferences across five Python linters.
    
    \item Our analysis reveals unique characteristics of each linter in four aspects, showing that no single linter excels in all aspects. 
   
   \item We offer practical suggestions to help linter users select and use linters, and outline directions for future research.

\end{itemize}

%% file: motivation.tex
\subsection{Python Linter Selection}~\label{linter_sel} 
\vspace{-1mm}
To ensure a comprehensive study of Python linters, we first use ``linter'' and ``python'' as keywords to search on GitHub. 
Among the top 300 returned projects, we identify 29 Python linters. 
However, analyzing all linters is impractical and unnecessary, as some have low star counts, few downloads, limited checks, or are integrated into others. 
For example, although isort is popular, it only sorts import statements, and linters like pycodestyle and Pyflakes have already been wrapped into Flake8. 
To select a representative set, we based our choices on two criteria: high popularity and broad checks. 
Ultimately, we select five linters: Ruff, Pylint, Flake8, Mypy and Black. 
Table~\ref{linter_overview} shows the GitHub star count, PyPI download numbers for 2024, and a brief description of each linter.  


\begin{table}[h]
\centering
\caption{Information of Studied Five Python Linters}
\label{linter_overview}
\resizebox{\textwidth}{!}{
\begin{tabular}{|l|c|c|l|}
\hline
\multicolumn{1}{|c|}{Linter} & \multicolumn{1}{c|}{Stars} & \multicolumn{1}{c|}{Downloads} & \multicolumn{1}{c|}{Description} \\
\hline
Ruff & 39.4k & 266.5M & 
A fast Python linter written in Rust.
\\
\hline
Pylint & 5.4k & 333.3M & 
\begin{tabular}[c]{@{}l@{}}
Check for errors, enforce a coding standard, and look for code smells.
\end{tabular}
\\
\hline
Flake8 & 3.6k & 405.8M & 
Check the style and quality of Python code.
\\
\hline
Mypy & 19.3k & 382.2M & 
Check  variables and functions in Python code.
\\
\hline
Black & 40.2k & 536.6M & 
The uncompromising Python code formatter. 
\\
\hline
\end{tabular}
}
\vspace{-5mm}
\end{table}

\vspace{-1mm}
\subsection{Python Linter in Developer Discussion} 
\vspace{-1mm}
To identify the most frequently discussed aspects of linters, we turn to public developer discussion websites like Stack Overflow~\cite{stack_overflow} and Reddit~\cite{reddit}. 
Reddit is included in our analysis because it hosts a wide range of developer communities, offering insights into both broad discussion and user experiences, which can complement more technical posts on Stack Overflow. 
We search for the names of the five selected linters as keywords, collecting the top 200 questions: 100 from Stack Overflow and 100 from Reddit. 
By analyzing these posts, we identify four key aspects of Python linters that users focus on, noting that a post may cover multiple aspects. 


\noindent \textbf{(1) Linting rule checks:} Due to lots of linters and linting rules, users are often confused about the specific checks a linter performs and the differences between their rule checks, accounting for 37\%. 
For example, in a Stack Overflow \href{https://stackoverflow.com/questions/78191580/how-to-catch-potentially-undefined-variables-with-pylint-or-mypy/78193468#78193468}{post}, a user mistakenly believed that Mypy and Pylint do not check whether an object is used before being defined. 
In reality, Mypy excels at object usage checks and includes the rule check. 
It indicates that users need a systematic understanding of linting rules across different linters. 

\noindent \textbf{(2) Linter speed:} Developers often praise specific linters for their speed or express dissatisfaction with slow ones, accounting for 21\%.  
However, when comparing linter speeds, they typically lack reliable evidence, relying instead on personal project experiences or anecdotal feedback. 
For example, in \href{https://www.reddit.com/r/Python/comments/10lsovt/ruff\_a\_new\_fast\_and\_correct\_python\_checkerlinter/}{a post} on Reddit, users praised Ruff for its speed, with some even switching from Flake8, but no specific running times or projects were provided. 
Although Ruff compares the speed of different linters using linters' default configuration on a project, relying on a single project for performance assessments is insufficient to draw reliable conclusions.

\noindent \textbf{(3) Linter configuration:} To meet the diverse needs of users, linters offer flexible configurations. 
However, the lack of a systematic understanding of these configurations often leads to user confusion regarding configuration methods and options, 
 accounting for 38\%. 
 For example, in \href{https://stackoverflow.com/questions/49220022/how-can-mypy-ignore-a-single-line-in-a-source-file}{a post} on Stack Overflow, 
  users were unsure how to configure Mypy to ignore a single line of code. 
   This can actually be done using the inline configuration method via comments.

\noindent \textbf{(4) Linter preference:}
Users frequently seek guidance on preferred linter usage for most projects, including selecting linters, configurations, and linting rules, accounting for 37\%. 
For example, in \href{https://www.reddit.com/r/Python/comments/15u6fey/what_are_your_best_practices_for_coding_in_python/}{a post} on Reddit, users ask about others' preferences on linters and best coding practices.

To address users' concerns on Python linters, we systematically analyze the following four research questions: 

\begin{enumerate}[fullwidth,itemindent=0em,leftmargin = 0pt]
\item[\textbf{RQ1:}] What linting rules are enforced by Python linters?
\item[\textbf{RQ2:}] What is the running time of Python linters on projects? 
\item[\textbf{RQ3:}] How are Python linters configured? 
\item[\textbf{RQ4:}] What are the preferences for selecting linters, configuring linters, and choosing linting rules across projects?
\end{enumerate}

%% file: result_four_rqs_no_features.tex

\subsection{RQ1: What linting rules are enforced by Python linters?}~\label{linting_rule}
\vspace{-2mm}
\subsubsection{Motivation} 
Understanding the rules enforced by each linter is essential for users to select suitable linters.

\subsubsection{Approach} \label{lint_rule_aprch}
We extract linting rules from linter official documentation~\cite{ruff_rule,pylint_rule,flake8_rule,mypy_rule, black_rule}, where each rule is clearly defined. 
In total, there are 1,551 linting rules: 863 for Ruff, 469 for Pylint, 127 for Flake8, 70 for Mypy and 22 for Black. 

We then assess linting rule overlap across linters. 
For every rule in a linter, two annotators independently determine whether it is semantically equivalent to any rule in other linters based on their documentation. 
The Cohen's kappa of 0.89 indicates substantial agreement between the annotators, and all remaining disagreements were resolved through discussion.

Finally, we classify these linting rules using a card sorting approach in two iterations~\cite{card_sorting}. 
In the first iteration, we randomly sample 309 linting rules with a confidence level of 95\% and a confidence interval of 5 from 1,551 linting rules~\cite{singh2013elements}. 
 Two annotators independently read the description of each rule and annotate its functionality with a brief summary. 
 They then discuss any disagreements in their summary and resolve them. 
After that, they work together to group the rules into categories with the corresponding definitions. 

In the second iteration, they independently classify the remaining linting rules based on the predefined categories. 
If a linting rule does not fit the predefined category, they provide a brief summary. 
It is found that no new categories are needed. 
The Cohen's kappa score for inter-annotator agreement is 0.78, indicating substantial agreement~\cite{kappa_value}. 
Finally, they discuss disagreements to reach an agreement.

\begin{figure}[h]
  \centering
  \includegraphics[width=3.2in]{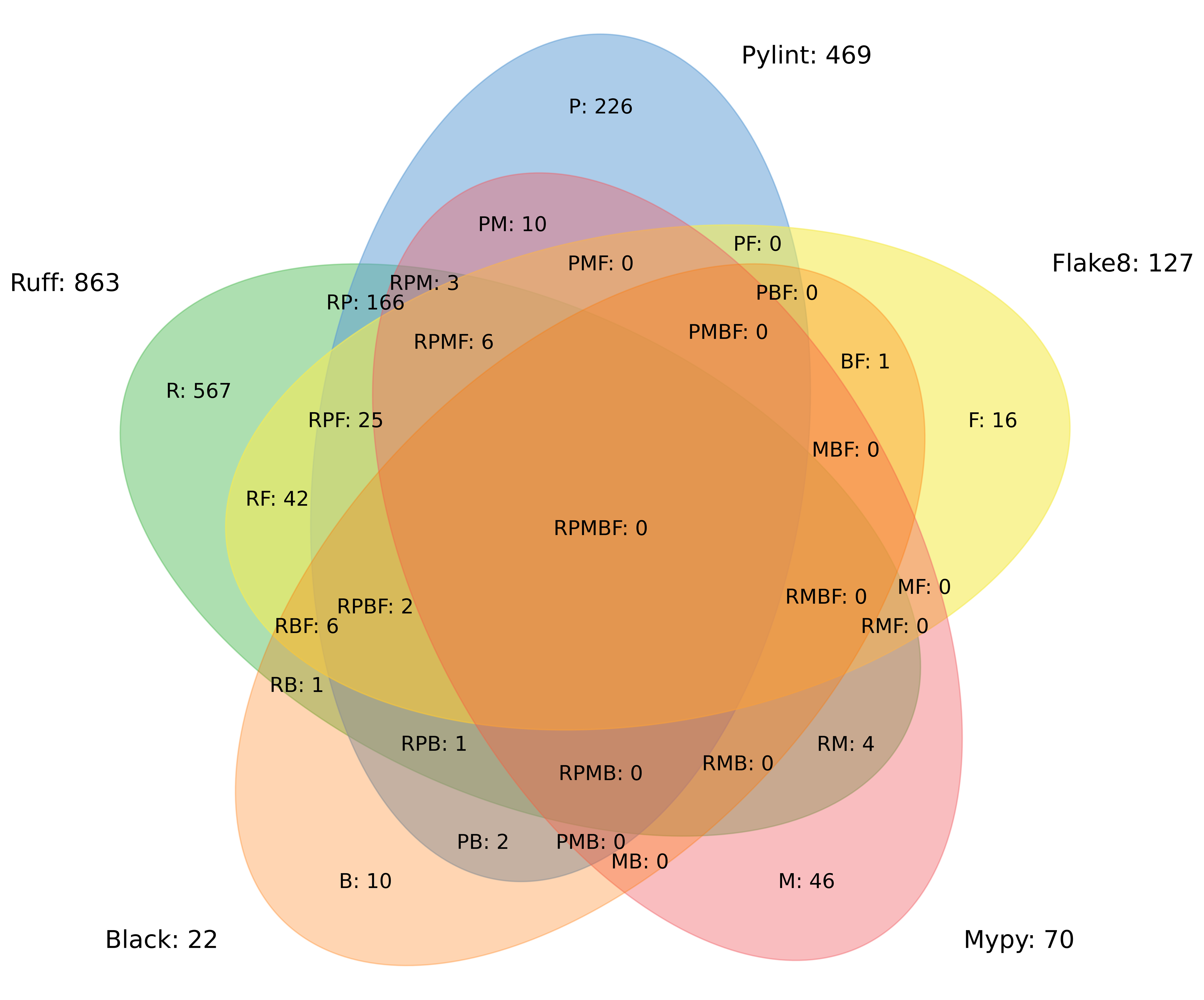}
  \setlength{\abovecaptionskip}{-1mm}
    \caption{Linting Rule Overlap Across Five Linters} 
    \label{fig:rule_overlap} 
\end{figure}

\subsubsection{Result} 
\noindent \textbf{(1) Linting Rule Overlap:}
Figure~\ref{fig:rule_overlap} shows the number of linting rules uniquely shared by specific linter combinations. 
Among the 1,551 linting rules across five linters, 865 rules (56\%) are exclusive to a single linter, and 686 rules (44\%) exist in at least two linters. 
There is no linting rule across the five linters. 
The results indicate that while linters often share many same checks, each linter also maintains its own unique checks. 
If users want to maximize rule coverage with fewer linters, using Ruff alone covers over 50\% of the linting rules. 
Combining Ruff and Pylint covers over 80\%, and using Ruff, Pylint, and Mypy together provides near-complete coverage (98\%). 
The rule overlap for each linter is as follows: 
\begin{itemize}[leftmargin=10pt]
\item  Ruff, with the largest number of rules (863), has 567 unique rules (66\%). 
Among the remaining rules, the number of overlaps with the other four linters is ranked as: Pylint \(>\) Flake8 \(>\) Mypy \(>\) Black.

\item  Pylint, with 469 rules, has 226 unique rules (48\%). 
The remaining 243 rules overlap with the other four linters, ranked as: Ruff \(>\) Flake8 \(>\) Mypy \(>\) Black.

\item  Flake8, with 127 rules, only has 16 unique rules (13\%), indicating most of its rules can be replaced by other linters. 
The remaining 111 rules overlap in the order of: Ruff \(>\) Pylint \(>\) Black \(>\) Mypy.

\item  Mypy, despite having only 70 rules, has the highest proportion of unique rules at 66\% (46 rules). 
The remaining 24 rules overlap with Pylint, Flake8 and Ruff, with the overlap ranked as: Pylint \(>\) Ruff \(>\) Flake8.

\item  Black, with the fewest rules (22), has 10 unique rules (45\%). 
The remaining 12 rules are covered by three linters, in the order of: Ruff = Flake8 \(>\) Pylint. 
Of these, 9 rules (41\%) are covered by Ruff and Flake8, 5 are covered by Pylint. 
\end{itemize}

\vspace{-2mm}
\begin{tcolorbox}[colframe=black, width=\linewidth,boxrule=1pt,  boxsep=2pt,left=1pt,right=1pt,top=1pt,bottom=1pt, sharp corners]
\textbf{Finding 1:} Across five Python linters, 56\% of the 1,551 rules are unique to a single linter and 44\% appear in two or more linters, with none common to all five. 
To maximize rule coverage with minimal redundancy, Ruff alone covers over half of all rules, Ruff and Pylint over 80\%, and adding Mypy provides nearly complete coverage (98\%). 
\end{tcolorbox}

\begin{table}[h]
\scriptsize
\centering
\caption{Linting Rule Category}
\label{rule_definition}
\begin{tabular}{|m{2.5cm}|m{5.5cm}|m{4.5cm}|}
\hline
\multicolumn{1}{|c|}{Category} & \multicolumn{1}{c|}{Definition} & \multicolumn{1}{c|}{Example} \\
\hline
  Wrong Syntax 
    & Code uses incorrect syntax that causes the code to fail to run or running behavior does not meet expectations. & 
\href{https://flake8.pycqa.org/en/latest/user/error-codes.html}{[Flake8: F706]} A return statement outside of a function/method.\\ \hline
  Wrong Usage of Object 
    & Code objects are incorrectly declared, implemented or
referenced. & \href{https://mypy.readthedocs.io/en/stable/error\_code\_list.html\#check-that-a-variable-is-not-used-before-it-s-defined-used-before-def}{[Mypy:used-before-def]} A variable is used before it is defined. \\  \hline
  Logic Contradiction
    & Code expressions or statements contain contradictory logic. & \href{https://pylint.readthedocs.io/en/stable/user_guide/messages/warning/modified-iterating-list.html}{[Pylint: W4701]} While a loop is iterating over a list, its block simultaneously adds or removes elements from the list. \\ 
\hline
  Style Convention
    & Code formatting follows stylistic conventions for users to better read code. & \href{https://black.readthedocs.io/en/stable/the\_black\_code\_style/current\_style.html\#empty-lines}{[Black: EmptyLine]} 
    Code does not have spurious vertical whitespace.
    \\  \hline
  Best Practice 
    & Code implementation follows optimal practices.  & \href{https://docs.astral.sh/ruff/rules/delete-full-slice/}{[Ruff: FURB131]} Use ``clear()'' instead of ``del'' to delete the entire slice of a list or dictionary for better performance. \\
\hline
\end{tabular}
\vspace{-0.4cm}
\end{table}

\noindent \textbf{(2) Linting Rule Category across Linters:} We define five linting rule categories and then analyze how each category is represented across the five linters.

\noindent \textbf{Linting rule category: }Table~\ref{rule_definition} presents the five categories of linting rules in five linters.

\begin{itemize}[leftmargin=10pt]
\item \textbf{Wrong syntax: }Linter checks whether the code uses incorrect syntax that may cause the code to fail to run or the running behavior does not conform to the expected behavior. 
Four linters (Ruff, Pylint, Flake8, and Mypy) perform such checks. 
For instance, F706 in Flake8 checks the return statement is incorrectly used outside of a function or method. 
As another example, W0711 in Pylint checks whether code wrongly uses ``or'' to catch multiple exceptions. 
While the code can normally run, the ``or'' operator doesn't work for catching multiple exceptions in a single except block. 

\item \textbf{Wrong usage of objects: }Linter checks whether code objects are incorrectly declared, implemented or referenced. 
Four linters (Ruff, Pylint, Flake8, and Mypy) check such code issues. 
For example, used-before-def in Mypy checks whether a variable is used before it's defined. 
For another example, E0301 in Pylint checks whether an ``\_\_iter\_\_'' method of a class is implemented without a corresponding ``\_\_next\_\_'' method. 

\item \textbf{Logic contradiction: }Linter checks whether code expressions or statements contain logical contradictions. 
Four linters (Ruff, Pylint, Flake8, and Mypy) check such code issues. 
For example, SIM220 in Ruff checks whether ``and'' expressions include both an expression and its negation. 
Such an error causes the expression to always evaluate to false. 
As another example, W4701 in Pylint checks if there are statements that add and remove elements from a list while iterating over the list in a loop statement. 
\item \textbf{Style convention: }Linter checks whether the code formatting affects readability. 
It mainly focuses on whitespace, naming conventions, and the comment conventions. 
All five linters perform such checks. 
For example, empty lines in Black checks whether there is spurious vertical whitespace.
For another example, E741 in Flake8 checks whether variables are named ``l'', ``O'', or ``I''.

\item \textbf{Best practice: }Linter checks whether the code functional implementation follows optimal practices. 
Four linters (Ruff, Pylint, Flake8, and Mypy) perform such checks. 
For example, FURB131 in Ruff checks whether code uses the `clear()' method to remove all items instead of `del' statements, as it offers better performance. 
For another example, E721 in Flake8 checks whether code wrongly uses `==' to compare types instead of `isinstance()'. 



\end{itemize}

\noindent \textbf{Category distribution by linters:}
Table~\ref{rule_distribution} shows the number of linting rules in each category for the five linters.
Among all 1,551 linting rules, most fall into the categories of \textit{best practice}, \textit{style convention} and \textit{wrong usage of objects}, which account for 47\%, 25\%, and 19\%, respectively. 
The remaining two categories (\textit{wrong syntax} and \textit{logical contradiction}) account for less than 10\% of all rules. 
The detailed rule category for each linter is as follows: 

\begin{itemize}[leftmargin=10pt]
\item Ruff covers all five rule categories. 
Compared to other linters, Ruff has the largest number of linting rules in \textit{logical contradiction}, \textit{style convention}, and \textit{best practice}.

\item Pylint also checks all five rule categories, particularly excelling in \textit{wrong syntax} and \textit{wrong usage of objects}, with the largest number of rules in these two categories compared to other linters.

\item Although Flake8 spans all five categories, it checks very few rules for \textit{logical contradiction}, with only two rules in this category. 

\item Mypy, although covering five categories, mainly focuses on \textit{wrong usage of objects} and \textit{style convention}, with fewer than 10 rules in each of the other three categories.

\item Black only covers a single category: \textit{style convention}. 
\end{itemize}

\begin{tcolorbox}[colframe=black, width=\linewidth,boxrule=1pt,  boxsep=2pt,left=1pt,right=1pt,top=1pt,bottom=1pt, sharp corners]
\textbf{Finding 2:} Across the five Python linters, linting rules have five categories: \textit{wrong syntax}, \textit{wrong usage of objects}, \textit{logic contradiction}, \textit{style convention} and \textit{best practice}, with nearly half of the rules focused on \textit{best practice}. 
Different linters show preferences for specific categories, suggesting that users can prioritize linters based on the types of issues they want to check in their code. 
Notably, Pylint leads the first two categories, and Ruff leads the last three.
\end{tcolorbox}

\begin{table}
\caption{Distribution of five rule categories across five linters}
\label{rule_distribution}
\centering
\begin{tabular}{|l|r|r|r|r|r|r|}
\hline
    \multicolumn{1}{|c|}{Rule Category} & \multicolumn{1}{c|}{Ruff} & \multicolumn{1}{c|}{Pylint} & \multicolumn{1}{c|}{Flake8} & \multicolumn{1}{c|}{Mypy} & \multicolumn{1}{c|}{Black} & \multicolumn{1}{c|}{Total}  \\
        \hline
        Wrong Syntax & 35 & 42 & 12 & 2 & 0 & 91\\
        \hline
        Wrong Usage of Objects  & 78 & 143 & 23 & 47 & 0 & 291 \\
        \hline
        Logic Contradiction & 35 & 22 & 2 & 1 & 0 & 60 \\
        \hline
        Style Convention & 213 & 65 & 71 & 12 & 22 & 383 \\
        \hline
        Best Practice & 502 & 197 & 19 & 8 & 0 & 726 \\
        \hline
        Total & 863 & 469 & 127 & 70 & 22 & 1,551\\
        \hline
\end{tabular}
\vspace{-0.3cm}
\end{table}

\subsection{RQ2: What is the running time of Python linters on projects?}~\label{linting_runtime}
\vspace{-3mm}
\subsubsection{Motivation} 
Linter running time is a key factor when choosing a linter, especially in CI workflows. 
Measuring and comparing execution time of linters helps users choose efficient linters and guides maintainers in optimizing performance. 

\subsubsection{Approach} 
When using linters, developers typically configure different linting rules based on project-specific requirements. 
However, due to the large number of linting rules, exhaustively evaluating all possible rule combinations is infeasible. 
To address this challenge, we measure running time under three setups: (i) the default rule setup representing typical running time, (ii) the all rules setup representing the upper bound on running time, and (iii) the single rule setup representing the lower bound on running time. 
All other configuration options remain at their default values. 

To balance evaluation cost while maximizing project coverage, we randomly select 1,000 Python projects from the 10,000 most-starred GitHub projects for the default-rule and all-rules setups. The single-rule setup requires a separate runtime measurement for every evaluated rule on each project, resulting in hundreds of rule–project measurements per project. To limit the computational cost, we randomly sample 278 projects from these 1,000 projects for the single-rule setup, using a 95\% confidence level and a 5\% margin of error~\cite{singh2013elements}.

For reliable running time measurement of a linter on a project, we follow prior studies~\cite{georges2007statistically,traini2021software,crape2020rigorous,laaber2019software,zhang2023faster} and execute each linter on the project multiple times on at least three virtual machine instances. 
For each virtual machine instance, we perform three warm-up runs followed by ten executions. 
If the running time uncertainty remains above 0.05, additional virtual machine instances are launched until the uncertainty falls below this threshold. 
Finally, we compute the average runtime across all executions as the measured runtime of the linter on the project.

After obtaining the running time of each linter on each project, we explore linters' running time differences. 
Specifically, we analyze: (i) whether five linters exhibit significant running time differences for each rule setup, (ii) whether different setups of each linter exhibit significant running time differences, (iii) whether different rules within a linter have significant running time differences, and (iv) whether the running time of a rule across different projects has high variability. 

Building on previous studies~\cite{sheng2020adversarial,qiu2019cross,park2023generative}, to assess significant differences in multiple groups, we first apply the Kruskal–Wallis H test~\cite{kruskal1952use}, where a p-value$<$0.05 indicates significant differences. 
If so, we further apply Scott–Knott ESD test~\cite{sk_esd} to divide the groups into statistically different clusters. 
This method is used in (i), (ii), and (iii). 
For instance, in (i), we first use Kruskal–Wallis H test to check if the running times of five linters differ significantly. 
If p-value$<$0.05, we apply Scott–Knott ESD test to divide the five linters into distinct groups based on their running times. 
To measure variability within a group, i.e., running times across multiple projects under the same rule, we compute the coefficient of variation (CV)~\cite{anda2008variability,reed2002use}, which is applied in (iv).  
If CV$>$1, it is generally considered to indicate high variability~\cite{anda2008variability}.

\begin{figure}[h]
  \centering
  \includegraphics[width=4in]{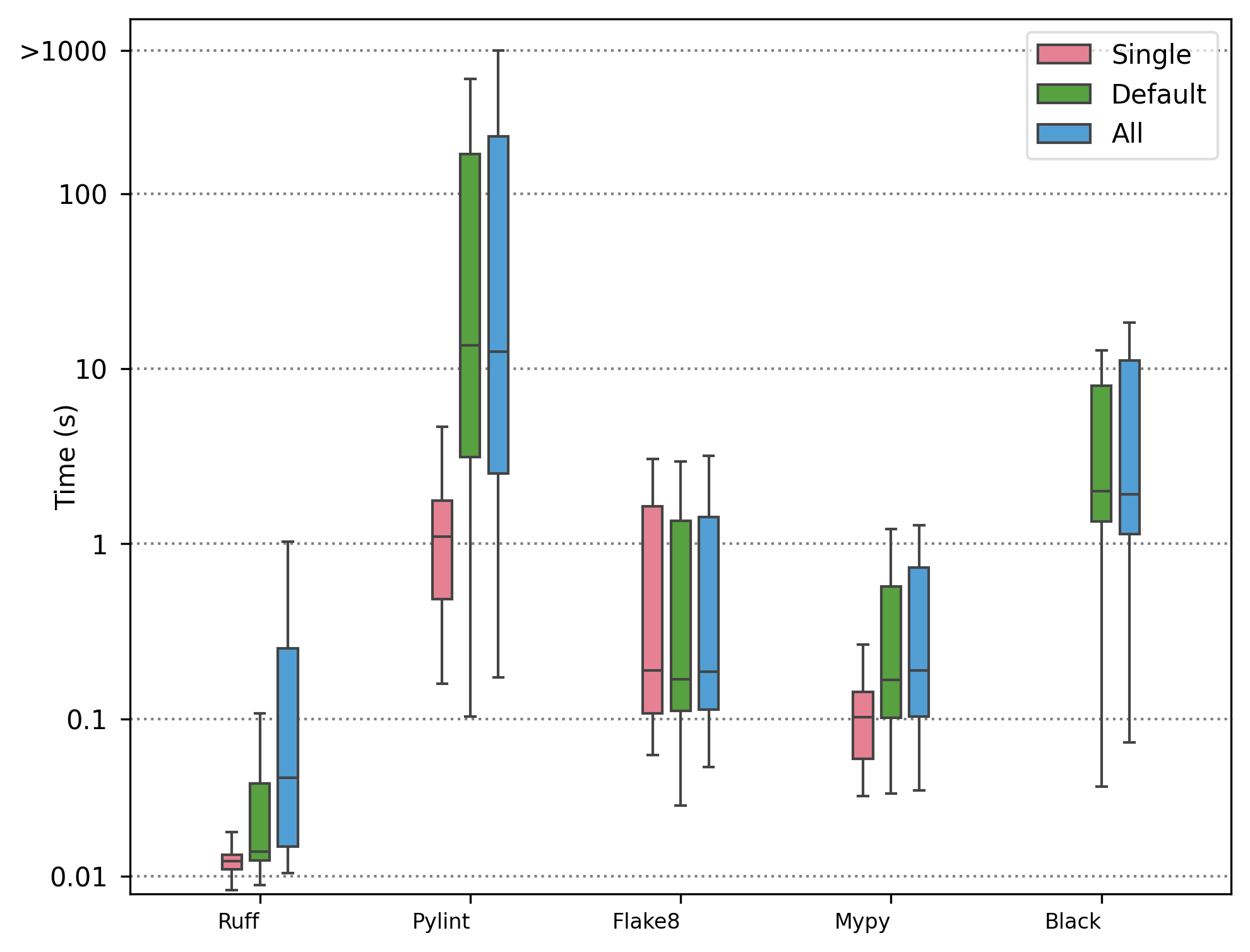}
  \setlength{\abovecaptionskip}{-1mm}
   \caption{The Runtime of Five Python Linters under Three Setups} 
    \label{fig:runtime_linter} 
    \vspace{-0.4cm}
\end{figure}

\subsubsection{Results}

\noindent \textbf{(1) Runtime Analysis of Linters:} 
Figure~\ref{fig:runtime_linter} shows the running time of the five linters under three rule setups. 
Table~\ref{tab:linter-runtimes} shows linters' running time differences. 

\begin{itemize}[leftmargin=10pt]
\item \textbf{Minimum running time of linters}: Enabling a single linting rule for each project reflects the linter's minimum running time. 
As shown in the single rule setup of Figure~\ref{fig:runtime_linter} and Table~\ref{tab:linter-runtimes}, except for Black, which does not support enabling a single linting rule, the speed ranking from fastest to slowest is: Ruff, Mypy, Flake8 and Pylint,  
operating within 5.5ms$\sim$23.1ms, 55.9ms$\sim$411ms, 79.2ms$\sim$5.4s, 281.4ms$\sim$7s, respectively. 



\item  \textbf{Typical running time of linters}: Enabling the default linting rules for each project reflects the linter's typical runtime. 
As shown in the default setup in Figure~\ref{fig:runtime_linter} and Table~\ref{tab:linter-runtimes}, the speed ranking from fastest to slowest is Ruff, Mypy, Flake8, Black and Pylint, 
operating 
within ranges of 5.1ms$\sim$129.2ms, 57.4ms$\sim$1.8s, 50.3ms$\sim$5.2s, 61.3ms$\sim$19.6s and 110.6ms$\sim$823.5s (14min), respectively. 
%

\begin{table}[h]
\centering
\caption{Linter Runtime Differences}
\label{tab:linter-runtimes}
\resizebox{\textwidth}{!}{
\begin{tabular}
{|c|c|c|c|c|c|c|}
\hline
\multirow{2}{*}{Linter} 
  & \multicolumn{3}{c|}{ 
  Linter Speed Rank
  } 
  & \multirow{2}{*}{
  Setup Runtime Rank
  } 
  & \multirow{2}{*}{
  Sig.
  }
  
  & \multirow{2}{*}{
  
  CV$>$1
  } \\
\cline{2-4}
& single rule & default ruleset & all rules& & & \\
\hline
Ruff   
  & \multirow{5}{*}{
  
  \begin{tabular}[t]{@{}l@{}}
  1 \\ 4\\3\\2\\--
  \end{tabular}
  }
  & \multirow{5}{*}{
    \begin{tabular}[t]{@{}l@{}}
    1 \\ 5\\3\\2\\4
    \end{tabular}
  }
  & \multirow{5}{*}{
  \begin{tabular}[t]{@{}c@{}}
  1 \\ 5\\3\\2\\4
  \end{tabular}
  }
  & single$<$default$<$all 
  & \checkmark 
  & 26.7\% \\
Pylint 
  &
  &
  &
  & single$<$default$\approx$all 
  & \checkmark 
  & 100\% \\
Flake8 
  &
  &
  &
  & single$\approx$default$\approx$all
  & 
  & 100\% \\
Mypy   
  &
  &
  &
  & single$<$default$\approx$all
  & \checkmark 
  & 100\% \\
Black  
  &
  &
  &
  & default$\approx$all
  & -- 
  & -- \\
\hline
\multicolumn{7}{@{}l@{}}{%
  \begin{tabular}{p{\linewidth}}
    \textit{Note:} single, default and all denote the linter's running time with a single rule enabled, with its default rule set, and with all rules enabled, respectively. 
    Sig. marks the significant running time difference across different rules. `--' means unavailable data. 
    CV$>$1 gives the percentage of rules whose running times vary widely across projects.
  \end{tabular}%
}
\end{tabular}
}
\end{table}

\item  \textbf{Maximum running time of linters}: Enabling all linting rules for each project reflects the linter's maximum running time.
As shown in the all rules setup in Figure~\ref{fig:runtime_linter} and Table~\ref{tab:linter-runtimes}, the speed ranking from fastest to slowest is Ruff, Mypy, Flake8, Black and Pylint, 
operating within ranges of 10.5ms$\sim$1s, 58.9ms$\sim$2s, 72.3ms$\sim$5.5s, 86.3ms$\sim$33.7s, and 314ms$\sim$$>$1000s ($>$17min), respectively. 

\item \textbf{Running time differences across three rule setups}: As shown in \textit{Setup Runtime Rank} column in Table~\ref{tab:linter-runtimes}, for Ruff, running times of all three rule setups exhibit significant differences, their speed rank is single $<$ default $<$ all. 
This indicates that their execution time scales with the number of rule checks. 
In contrast, for Pylint and Mypy, their speed ranking of three setups are both single $<$ default $\approx$ all, indicating that running time of the two linters may be constrained by factors other than the number of rules, such as the computational overhead of specific rule checks. 
For Flake8 and Black (which does not support single-rule setup), their running times of all rule setups do not exhibit significant differences, indicating their running times basically are unaffected by the number of rule checks. 




\item \textbf{Runtime differences across linting rules}: As shown in \textit{Sig.} column in Table~\ref{tab:linter-runtimes}, under the single rule setup, running times for different linting rules exhibit significant differences for Ruff, Pylint and Mypy. 
This result implies that Ruff, Pylint, and Mypy incur different running times for specific checks, enabling performance gains when users enable only less expensive rules. 
In contrast, for Flake8, the running times for different linting rules do not exhibit significant differences, and therefore cannot benefit from this optimization. 
Particularly, since Black does not support single rule configuration, the running time of individual linting rules is unavailable.

\item \textbf{Running time variability of a linting rule across projects}: As shown in \textit{CV$>$1} column in Table~\ref{tab:linter-runtimes}, all rules in Pylint, Flake8, and Mypy exhibit high variability under different projects, whereas only 26.7\% of Ruff's rules do so. Particularly, Black does not support the single rule configuration.
For Pylint, Flake8, and Mypy, even setting the same linting rule, its running time is highly project-specific. 
By contrast, most Ruff rules run with stable performance across diverse projects. 



\end{itemize}

\vspace{-2mm}
\begin{tcolorbox}[colframe=black, width=\linewidth,boxrule=1pt,  boxsep=2pt,left=1pt,right=1pt,top=1pt,bottom=1pt, sharp corners]
\textbf{Finding 3:} For the five linters, the speed ranking from fastest to slowest is as follows: Ruff, Mypy, Flake8, Black and Pylint, 
with running times of 5.1ms$\sim$1s, 55.9ms$\sim$2s, 50.3ms$\sim$5.5s, 61.3ms$\sim$33.7s, and 110.6ms$\sim$$>$1000s.  
Except for Flake8 and Black, the running times of Ruff, Pylint and Mypy are affected by different rule checks, indicating that users can optimize their performance by rule selection. 


\end{tcolorbox}

\subsection{RQ3: How are Python linters configured?}\label{config} 
\subsubsection{Motivation} 
With the increasing variety of configuration mechanisms, we aim to investigate which configuration methods and options are supported by Python linters. 

\subsubsection{Approach} 

We begin by summarizing the configuration methods supported by each linter based on their official documentation~\cite{ruff_config,pylint_config,flake8_config,mypy_config,black_config}. 
We then collect all configuration options across all configuration methods and categorize them using a card sorting approach, following the same procedure used for classifying linting rules in Section~\ref{lint_rule_aprch}. 




\subsubsection{Result} We summarize configuration methods and configuration options of the five linters as follows: 

\begin{table}[h]
  \centering
  \caption{Configuration Methods supported by five linters}
  \label{config_method}
  \begin{tabular}{|l|c|c|c|c|c|c|}
    \hline
    \multirow{3}{*}{Tool}
    & \multirow{3}{*}{\#CL}
    & \multicolumn{4}{c|}{Configuration File}
    & \multirow{3}{*}{\#IC} \\
    \cline{3-6}
    &  & \multicolumn{3}{c|}{Project Configuration File} & \multirow{2}{*}{\#DC} & \\
    \cline{3-5}
    &  & pyproject.toml & setup.cfg & tox.ini &  & \\
    \hline
    Ruff   & \checkmark & \checkmark &  &   & \checkmark & \checkmark \\
    \hline
    Pylint & \checkmark & \checkmark & \checkmark & \checkmark & \checkmark & \checkmark \\
    \hline
    Flake8 & \checkmark &  & \checkmark & \checkmark & \checkmark & \checkmark \\
    \hline
    Mypy   & \checkmark & \checkmark & \checkmark &  & \checkmark  & \checkmark \\
    \hline
    Black  & \checkmark & \checkmark &  &  & \checkmark & \checkmark \\
    \hline
    \multicolumn{7}{@{}l@{}}{\begin{tabular}{@{}l@{}} Note: \#CL is command line configuration; \#DC is dedicated configuration \\file; \#IC is inline configuration. \end{tabular}}\\
  \end{tabular}
 \vspace{-4mm}
\end{table}

\noindent \textbf{(1) Configuration Method}: 
Table~\ref{config_method} shows the configuration methods supported by five Python linters. 
There are four configuration methods: 
command line configuration, project configuration files, dedicated configuration files, and inline configuration. 
Project configuration files include three specific formats: \textit{pyproject.toml}, \textit{setup.cfg}, and \textit{tox.ini}. 
The five linters support all four configuration methods. 
Notably, Pylint is the only linter that supports both four configuration methods and three project-level configuration formats. 
The configuration methods are as follows: 
\begin{itemize}[leftmargin=10pt]
\item  \textbf{Command line configuration}: Linter configurations are specified directly via command-line arguments during execution. 
It is suitable for one-off runs or use in continuous integration scripts, but it is difficult to maintain and not suitable for complex or long-term configurations. 
All five Python linters support the command line configuration.

\item \textbf{Project configuration files}: Linter configurations are defined in project configuration files such as \textit{pyproject.toml}, \textit{setup.cfg} and \textit{tox.ini}. 
This configuration method streamlines tool configuration, minimizes scattered files, and enhances cross-tool compatibility. 
With PEP 518~\cite{pep518}, \textit{pyproject.toml} has become the standard for configuring Python packages, offering a one-stop shop for most project configuration needs. 
\textit{setup.cfg}, a legacy format in the Python packaging ecosystem, remains widely used by projects~\cite{setup_cfg}.
\textit{tox.ini} specifies the setup and execution of various tools like test runners and linters~\cite{tox_ini}. 
 Among the five linters, only Pylint supports all three configuration formats, Mypy and Flake8 support two configuration formats (\textit{pyproject.toml} and \textit{setup.cfg} for Mypy, \textit{setup.cfg} and \textit{tox.ini} for Flake8), while Ruff and Black only support \textit{pyproject.toml}. 
Overall, the most widely supported project configuration formats are ranked as follows: \textit{pyproject.toml} \(>\) \textit{setup.cfg} \(>\) \textit{tox.ini}. 

\item  \textbf{Dedicated configuration files}: Linter configurations are defined in standalone tool-specific files. 
The configuration method isolates linter settings from other project configurations, improving modularity and clarity. 
However, using separate configuration files for different tools results in fragmentation and increased maintenance overhead due to the lack of uniformity between tools. 
All the five Python linters support the dedicated configuration file method.

\item \textbf{Inline configuration}: The configuration is embedded directly into the source code through comments. 
Inline configuration provides fine-grained control over linter behavior, but excessive use can clutter the codebase and make maintenance hard. 
All five Python linters support the inline configuration.  
\end{itemize}

\vspace{-2mm}
\begin{tcolorbox}[colframe=black, width=\linewidth,boxrule=1pt,  boxsep=2pt,left=1pt,right=1pt,top=1pt,bottom=1pt, sharp corners]
\textbf{Finding 4:} All five Python linters support four configuration methods: \textit{command line configuration}, \textit{project configuration files}, \textit{dedicated configuration files}, and \textit{inline configuration}, but their support for project configuration formats differs. 
Only Pylint supports all three formats (\textit{pyproject.toml}, \textit{setup.cfg}, and \textit{tox.ini}), while Mypy and Flake8 support two, and Ruff and Black support only \textit{pyproject.toml}. 
Notably, the fact that four out of five linters support \textit{pyproject.toml} underscores a widespread convergence on the PEP 518 standard for centralized, cross-tool configuration. 
\end{tcolorbox}

\begin{table}[h]
    \centering
     \caption{The Distribution of Four Types of Configuration Options  
    }
    \label{tab:config_options}
    \scalebox{0.99}{
    \begin{tabular}{|l|c|c|c|c|c|c|}
        \hline
        \multicolumn{1}{|c|}{Option Type} & \multicolumn{1}{c|}{Ruff}  & \multicolumn{1}{c|}{Pylint} & \multicolumn{1}{c|}{Flake8} & \multicolumn{1}{c|}{Mypy} & \multicolumn{1}{c|}{Black} & \multicolumn{1}{c|}{Total} \\
        \hline
        Environment Setup & 7 & 7 & 2 & 11 & 4 & 31  \\
        \hline
        Scope Control     & 12& 9 & 4 & 10 & 9 & 44 \\
        \hline
        Rule Specification&135&130& 13& 39 & 8 & 325 \\
        \hline
        Result Handling   & 2 & 7 & 6 & 24 & 4 & 43 \\
        \hline
        Total             &156&153& 25& 84 & 25& 443 \\
        \hline
    \end{tabular}
    }
\end{table}

\noindent \textbf{(2) Configuration Option}: Table~\ref{tab:config_options} shows the distribution of four types of configuration options for each linter. 
The five linters have a total of 443 configuration options, divided into four categories: \textit{environment setup}, \textit{scope control}, \textit{rule specification}, and \textit{result handling}. 
The number of configuration options generally reflects a tool's flexibility: Ruff and Pylint each have over 100 options, with 156 and 153 options, respectively; Mypy has over 50 options (84 options); Flake8 and Black offer fewer than 50 options, each with 25 options. 


\begin{itemize} [leftmargin=10pt]
\item  \textbf{Environment setup}: Options control how the linter runs, including its execution mode, resource usage, and caching behavior. 
There are 31 such options, accounting for 7\% of all options.
Five linters are ordered by the number of options as follows: Mypy \(>\) Pylint = Ruff \(>\) Black \(>\) Flake8. 
For example, the ``jobs'' option in Pylint, which is to use multiple processes to speed up execution. 


\item  \textbf{Scope control}: Options that define the files or code regions the linter checks.  
There are 44 such options, accounting for 10\% of all options. 
Five linters are ordered by the number of options as follows: Ruff \(>\) Mypy \(>\) Pylint = Black \(>\) Flake8. 
For example, the ``exclude'' option in Ruff is used to skip a list of files during lint checks. 

\item  \textbf{Rule specification}:  Options that define which code issues the linter checks and how to solve these code issues. 
There are 325 such options, accounting for 73\% of all options. 
Five linters are ordered by the number of options as follows: Ruff \(>\) Pylint \(>\) Mypy \(>\) Flake8 \(>\) Black.  
For example, ``select'' in Flake8 is to specify rule sets linter checks. 

\item  \textbf{Result handling}:  Options that determine how linting results are formatted, logged, or finalized. 
There are 43 such options, accounting for 10\% of all options. 
Five linters are ordered by the number of options as follows: Mypy \(>\) Pylint \(>\) Flake8 \(>\) Black \(>\) Ruff.  
For example, ``color'' in Black is used to colorize the diff. 


\end{itemize}

\vspace{-2mm}
\begin{tcolorbox}[colframe=black, width=\linewidth,boxrule=1pt,  boxsep=2pt,left=1pt,right=1pt,top=1pt,bottom=1pt, sharp corners]
\textbf{Finding 5:} All five Python linters support four types of configuration options: \textit{environment setup}, \textit{scope control}, \textit{rule specification}, and \textit{result handling}, but their emphasis varies.  
\textit{Rule specification} is richest in Ruff and Pylint, each offering at least 130 options. 
\textit{Scope control} sees the most flexibility in Ruff and Mypy, each offering at least 10 options.
Mypy leads both \textit{environment setup} and \textit{result handling}.
For the other two linters, Flake8 and Black each provide only 25 options overall, indicating a more opinionated, minimal configuration flexibility. 
\end{tcolorbox}

\subsection{RQ4: What are the preferences for selecting linters, configuring linters, and choosing linting rules across projects?}\label{config_prefer}
\subsubsection{Motivation} Section~\ref{linting_rule} and Section~\ref{config} reveal the diversity of linting rules and the flexibility of the linter configurations. 
It is important to understand the projects' linter preferences to guide developers in using linters better. 
\subsubsection{Approach} 
To understand linter configuration preferences across projects, we first collect top 10,000 most-starred GitHub Python projects.  
For each linter, we first examine whether a project configures a linter, and then compute the number of projects using each configuration method. 
For the five linters, we identify a total of 3,094 projects. 
Finally, we analyze the five most frequently configured options, as well as the enabled and disabled linting rules across projects.

\subsubsection{Result} 
Among the 3,094 projects using the five Python linters, \textbf{the popularity of the five linters used in the projects is as follows: Flake8 \(>\) Ruff \(>\) Mypy \(>\) Black \(>\) Pylint,} accounting for 49.2\%, 36.7\%, 35.1\%, 24.3\%, and 21.0\%, respectively.
56.6\% of projects use more than one linter, indicating that many projects rely on multiple linters for comprehensive rule check. 
The linter preferences regarding configuration methods, configuration options, and linting rules are as follows:


\begin{figure}[h]
  \centering
  \includegraphics[width=4in]
  {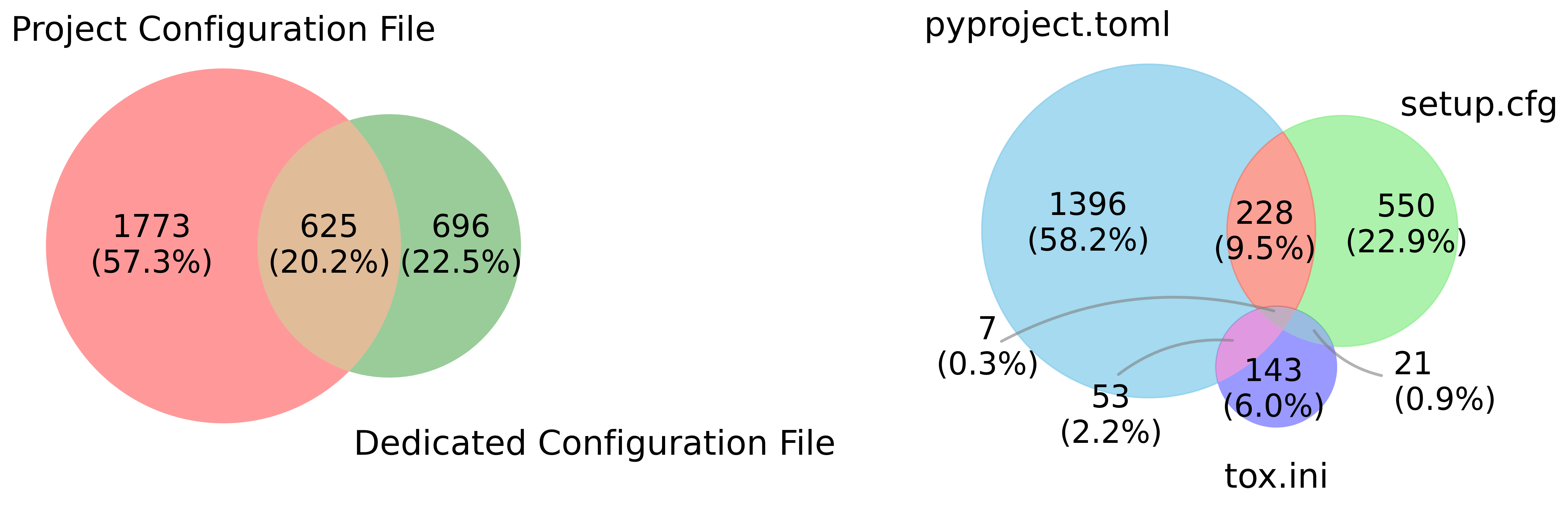}
    \caption{Distribution of Configuration Method Preference Across Five Linters} 
    \label{fig:all_config_file} 
\end{figure}

\noindent \textbf{(1) Configuration Method Preference across Projects}: Figure~\ref{fig:all_config_file} shows the overall distribution of configuration methods and configuration formats across all projects that use Ruff, Pylint, Mypy, Flake8 or Black. 

\noindent \textbf{Preference for project configuration vs. dedicated files}: As shown in the left panel of Figure~\ref{fig:all_config_file}, among these projects, 57.3\% configure linters using only project configuration files, 22.5\% rely solely on dedicated configuration files, and 20.2\% use two configuration methods. 
This distribution highlights projects’ inclination to centralize  linter settings in a single, standardized project configuration file rather than in separate, tool-specific configuration files. 

\noindent \textbf{Preference for project configuration formats}:  
As shown in the right panel of Figure~\ref{fig:all_config_file}, when projects configure linters via project configuration files, the popularity of formats follows the order: \textit{pyproject.toml} \(>\) \textit{setup.cfg} \(>\) \textit{tox.ini}. 
This trend indicates that many projects consolidate around the PEP
518 style: \textit{pyproject.toml}. 
Notably, 12.6\% of projects employ more than one project-level format, most commonly pairing \textit{pyproject.toml} with \textit{setup.cfg} (9.5\%). 
The mixed usage of linter configuration formats may highlight the transitional workflows: projects are likely migrating from legacy \textit{setup.cfg} to \textit{pyproject.toml} while maintaining backward compatibility. 

\begin{figure}[h]
  \centering
  \includegraphics[width=4.5in]{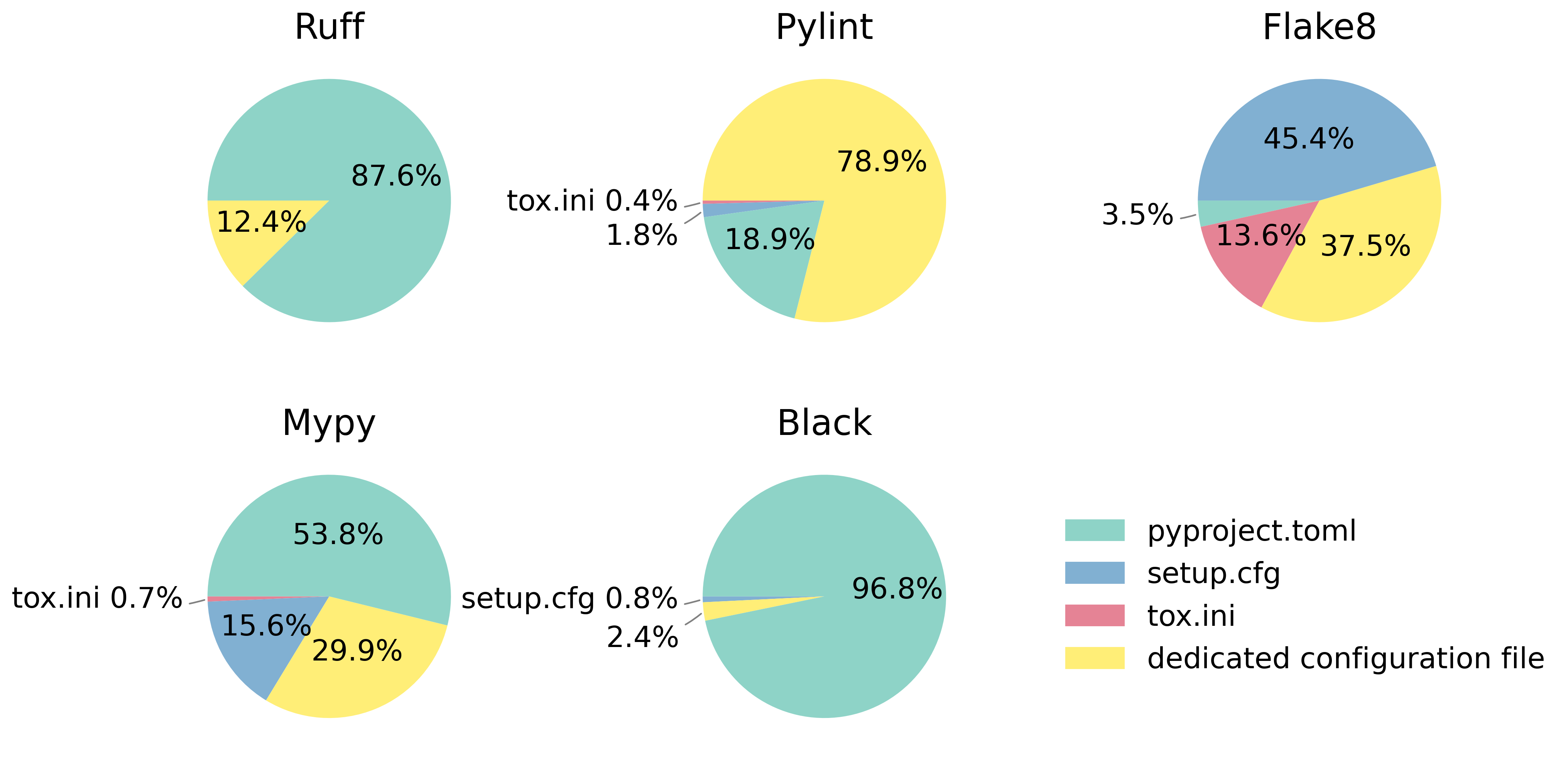}
\setlength{\abovecaptionskip}{-0.05cm}
    \caption{Distribution of Configuration Method Preference of Each Linter}
    \label{fig:tool_config_file} 
    \vspace{-0.4cm}
\end{figure}

\noindent \textbf{Configuration method preference for each linter}: 
Figure~\ref{fig:tool_config_file} illustrates the distribution of different configuration methods across projects for each Python linter. 
\begin{itemize}[leftmargin=10pt]

   \item For \textit{Ruff}, \textit{Mypy} and \textit{Black}, the majority of projects ($>$70\%) use project  configuration. 
   Furthermore, projects using these linters strongly prefer the \textit{pyproject.toml} format, which all account for more than 50\% of configurations. 

   \item For \textit{Flake8}, most projects ($>$60\%) also use project configuration,  with \textit{setup.cfg}, \textit{tox.ini} and \textit{pyproject.toml}, accounting for 45.4\%, 13.6\% and 3.5\%, respectively. 
   Notably, although \textit{Flake8} does not natively support \textit{pyproject.toml}, 3.5\% of projects use third-party plugins to enable configuration in \textit{pyproject.toml}. 
   The result further verifies projects'  preference for a single, centralized configuration file and how the plugin ecosystem compensates for gaps in native tool support. 

   \item For \textit{Pylint}, we see the opposite trend, with the majority of projects ($>$70\%) relying on separate, tool-specific configuration files. 
Nevertheless, among the minority that do adopt project configuration for Pylint, \textit{pyproject.toml} remains the predominant format (close to 20\%).
\end{itemize}

\vspace{-2mm}
\begin{tcolorbox}[colframe=black, width=\linewidth,boxrule=1pt,  boxsep=2pt,left=1pt,right=1pt,top=1pt,bottom=1pt, sharp corners]
\textbf{Finding 6:} 
Most projects using Ruff, Mypy, and Black use \textit{pyproject.toml}, while those using Flake8 prefer \textit{setup.cfg}. 
 By contrast, projects using Pylint predominantly rely on a dedicated tool-specific file. 
 While the PEP 518 (using \textit{pyproject.toml}) is widely accepted for its standardization and compatibility across tools, the fact that projects using different linters adopt distinct configuration methods reflects the varied preferences of developers based on the specificity and opinionated nature of the linters they use.
\end{tcolorbox}

\begin{table*}
  \centering
  \caption{Top Five Most Frequently Configured Options, enabled and disabled Linting Rules in Five Linters}
  \label{tab:option_freq}
  
\resizebox{\textwidth}{!}{
\begin{tabular}{|c|c|l|l|l|}
    \hline
    \multicolumn{1}{|c|}{Linter} & \multicolumn{1}{c|}{\#Proj} & \multicolumn{1}{c|}{Top-5 Used Options} & \multicolumn{1}{c|}{Top-5 Enabled Linting Rules} & \multicolumn{1}{c|}{Top-5 Disabled Linting Rules} \\
    \hline
    Ruff
      & 1135
      & \begin{tabular}[t]{@{}l@{}}
          \href{https://docs.astral.sh/ruff/settings/\#line-length}{line-length}: 68.1\% -- Rule Specification \\
          \href{https://docs.astral.sh/ruff/settings/\#lint_ignore}{ignore}: 59.4\% - Rule Specification \\
          \href{https://docs.astral.sh/ruff/settings/\#lint_select}{select}: 57.9\% - Rule Specification \\
          \href{https://docs.astral.sh/ruff/settings/\#target-version}{target-version}: 47.6\% - Environment Setup
          \\
          \href{https://docs.astral.sh/ruff/settings/\#lint_per-file-ignores}{per-file-ignores}: 36.4\% - Scope Control\\
          Total: 93.7\%
        \end{tabular}
      & \begin{tabular}[t]{@{}l@{}}
          \href{https://docs.astral.sh/ruff/rules/\#pyflakes-f}{F}: 64.4\% - Wrong Usage of Objects, \\ \ \ \ \ \ \ \ \ \ \ \ \ \ \ Best Practice and Wrong Syntax \\
          \href{https://docs.astral.sh/ruff/rules/\#error-e}{E}: 57.4\% - Style Convention \\
          \href{https://docs.astral.sh/ruff/rules/\#isort-i}{I}: 55.5\% - Style Convention and\\\ \ \ \ \ \ \ \ \ \ \ \ \ Wrong Usage of Objects\\
          \href{https://docs.astral.sh/ruff/rules/\#warning-w}{W}: 46.1\% - Style Convention and\\\ \ \ \ \ \ \ \ \ \ \ \ \ \ Wrong Syntax\\
          \href{https://docs.astral.sh/ruff/rules/\#pyupgrade-up}{UP}: 37.2\% - Best Practice and\\\ \ \ \ \ \ \ \ \ \ \ \ \ \ \ \  Style Convention\\
          Total: 75.2\%
        \end{tabular}
      & \begin{tabular}[t]{@{}l@{}}
          \href{https://docs.astral.sh/ruff/rules/line-too-long/}{E501}: 34.5\% - Style Convention \\
          \href{https://docs.astral.sh/ruff/rules/ambiguous-variable-name/}{E741}: 12.8\% - Style Convention  \\
          \href{https://docs.astral.sh/ruff/rules/lambda-assignment/}{E731}: 12.2\% - Best Practice \\
          \href{https://docs.astral.sh/ruff/rules/complex-structure/}{C901}: 10.4\% - Style Convention \\
          \href{https://docs.astral.sh/ruff/rules/module-import-not-at-top-of-file/}{E402}: 10.2\% - Style Convention \\
          Total: 48.3\%
        \end{tabular}\\
    \hline
    Pylint
    & 651
      & \begin{tabular}[t]{@{}l@{}}
          \href{https://pylint.readthedocs.io/en/stable/user_guide/configuration/all-options.html\#disable}{disable}: 76.0\% - Rule Specification \\
          \href{https://pylint.readthedocs.io/en/stable/user_guide/configuration/all-options.html\#max-line-length}{max-line-length}: 64.1\% - Rule Specification \\
          \href{https://pylint.readthedocs.io/en/stable/user_guide/configuration/all-options.html\#good-names}{good-names}: 56.4\% - Rule Specification \\
          \href{https://pylint.readthedocs.io/en/stable/user_guide/configuration/all-options.html\#reports}{reports}: 55.2\% - Result Handling \\
          \href{https://pylint.readthedocs.io/en/stable/user_guide/configuration/all-options.html\#min-similarity-lines}{min-similarity-lines}: 53.9\%  - Rule Specification\\
          Total: 84.3\%
        \end{tabular}
      & \begin{tabular}[t]{@{}l@{}}
          \href{https://pylint.pycqa.org/en/1.9/technical_reference/features.html\#id1} {old-raise-syntax}: 5.4\% - Wrong Syntax\\
          \href{https://pylint.pycqa.org/en/1.9/technical_reference/features.html}{indexing-exception}: 5.1\% - Wrong Syntax \\ 
          \href{https://pylint.readthedocs.io/en/stable/user_guide/messages/information/useless-suppression.html}{useless-suppression}: 3.2\% - Best Practice\\
          \href{https://pylint.readthedocs.io/en/stable/user_guide/messages/warning/unused-import.html}{unused-import}: 3.1\% - Best Practice \\ 
          \href{https://pylint.readthedocs.io/en/stable/user_guide/messages/warning/dangerous-default-value.html}{dangerous-default-value}: 2.6\%  - Wrong Usage \\ \ \ \ \ \ \ \ \ \ \ \ \ \ \ \ \ \ \ \ \ \ \ \ \ \ \ \ \ \ \ \ \ \ \ \ \ \ \ \  of Objects \\
          Total (10.5\%)
        \end{tabular}
      & \begin{tabular}[t]{@{}l@{}}          
          \href{https://pylint.readthedocs.io/en/stable/user_guide/messages/information/locally-disabled.html}{locally-disabled}: 36.6\% - Best Practice \\
          \href{https://pylint.readthedocs.io/en/stable/user_guide/messages/refactor/too-few-public-methods.html}{too-few-public-methods}: 36.3\% - Best Practice \\
          \href{https://pylint.readthedocs.io/en/stable/user_guide/messages/warning/fixme.html}{fixme}: 35.2\% - Style Convention \\
          \href{https://pylint.readthedocs.io/en/stable/user_guide/messages/refactor/too-many-arguments.html}{too-many-arguments}: 35.2\% - Best Practice\\
          \href{https://pylint.readthedocs.io/en/stable/user_guide/messages/convention/invalid-name.html}{invalid-name}: 30.1\% - Style Convention \\
          Total: 62.7\%
        \end{tabular}\\
    \hline
    Flake8
    & 1523
      & \begin{tabular}[t]{@{}l@{}}
          \href{https://flake8.pycqa.org/en/latest/user/options.html\#cmdoption-flake8-max-line-length}{max-line-length}: 76.3\% - Rule Specification \\
          \href{https://flake8.pycqa.org/en/latest/user/options.html\#cmdoption-flake8-ignore}{ignore}: 56.9\% - Rule Specification \\
          \href{https://flake8.pycqa.org/en/latest/user/options.html\#cmdoption-flake8-exclude}{exclude}: 55.6\% - Scope Control\\
          \href{https://flake8.pycqa.org/en/latest/user/options.html\#cmdoption-flake8-per-file-ignores}{per-file-ignores}: 21.2\% - Scope Control \\
          \href{https://flake8.pycqa.org/en/latest/user/options.html\#cmdoption-flake8-max-complexity}{max-complexity}: 18.4\% - Rule Specification \\
          Total: 98.6\%
        \end{tabular}
      & \begin{tabular}[t]{@{}l@{}}
          \href{https://flake8.pycqa.org/en/latest/user/error-codes.html}{F}: 15.4\% - Wrong Usage of Objects, \\ \ \ \ \ \ \ \ \ \ \ \ \ \ \ Best Practice and Wrong Syntax \\
          \href{https://pycodestyle.pycqa.org/en/latest/intro.html\#error-codes}{W}: 15.1\% - Style Convention and\\\ \ \ \ \ \ \ \ \ \ \ \ \ \ \ Wrong Syntax \\
          \href{https://pycodestyle.pycqa.org/en/latest/intro.html\#error-codes}{E}: 15.0\%- Style Convention\\
          \href{https://flake8.pycqa.org/en/latest/user/error-codes.html}{C}: 12.7\% - Best Practice \\
          \href{https://github.com/PyCQA/flake8-bugbear}{B}: 12.3\%- {Best Practice}\\
          Total: 15.6\%
        \end{tabular}
      & \begin{tabular}[t]{@{}l@{}}
          \href{https://pycodestyle.pycqa.org/en/latest/intro.html\#error-codes}{W503}: 49.1\% - Style Convention\\
          \href{https://pycodestyle.pycqa.org/en/latest/intro.html\#error-codes}{E203}: 39.3\% - Style Convention \\
          \href{https://pycodestyle.pycqa.org/en/latest/intro.html\#error-codes}{E501}: 33.0\% - Style Convention \\
          \href{https://pycodestyle.pycqa.org/en/latest/intro.html\#error-codes}{W504}: 17.9\% - Style Convention \\
          \href{https://pycodestyle.pycqa.org/en/latest/intro.html\#error-codes}{E402}: 13.1\% - Style Convention \\
          Total: 62.9\%
        \end{tabular}
       \\
    \hline
    Mypy
    & 1086
      & \begin{tabular}[t]{@{}l@{}}
          \href{https://mypy.readthedocs.io/en/stable/config_file.html\#confval-ignore_missing_imports}{ignore\_missing\_imports}: 49.2\% - Rule Specification \\
          \href{https://mypy.readthedocs.io/en/stable/config_file.html\#confval-warn_unused_ignores}{warn\_unused\_ignores}: 35.6\% - Rule Specification \\
          \href{https://mypy.readthedocs.io/en/stable/config_file.html\#confval-warn_redundant_casts}{warn\_redundant\_casts}: 32.1\% - Rule Specification \\
          \href{https://mypy.readthedocs.io/en/stable/config_file.html\#confval-python_version}{python\_version}: 30.9\% - Environment Setup \\
          \href{https://mypy.readthedocs.io/en/stable/config_file.html\#confval-check_untyped_defs}{check\_untyped\_defs}: 29.9\% - Rule Specification \\
          Total: 78.5\%
        \end{tabular}
      & \begin{tabular}[t]{@{}l@{}}
          \href{https://mypy.readthedocs.io/en/stable/error_code_list2.html\#check-that-type-ignore-include-an-error-code-ignore-without-code}{ignore-without-code}: 4.8\% - Best Practice\\
          \href{https://mypy.readthedocs.io/en/stable/error_code_list2.html\#check-that-expression-is-not-implicitly-true-in-boolean-context-truthy-bool}{truthy-bool}: 3.5\%- Wrong Usage of Objects \\
          \href{https://mypy.readthedocs.io/en/stable/error_code_list2.html\#check-that-expression-is-redundant-redundant-expr}{redundant-expr}: 3.1\% - Best Practice \\
          \href{https://mypy.readthedocs.io/en/stable/error_code_list2.html\#check-that-iterable-is-not-implicitly-true-in-boolean-context-truthy-iterable}{truthy-iterable}: 1.6\% - Wrong Usage of Objects \\
          \href{https://mypy.readthedocs.io/en/stable/error_code_list2.html\#check-that-methods-do-not-have-redundant-self-annotations-redundant-self}{redundant-self}: 1.5\% - Best Practice\\
          Total: 5.0\%
        \end{tabular}
      & \begin{tabular}[t]{@{}l@{}}
          \href{https://mypy.readthedocs.io/en/stable/error_code_list.html\#check-that-attribute-exists-attr-defined}{attr-defined}: 0.9\% - Wrong Usage of Objects \\
          \href{https://mypy.readthedocs.io/en/stable/error_code_list.html\#check-that-import-target-can-be-found-import-untyped}{import-untyped}: 0.9\% - Style Convention \\
          \href{https://mypy.readthedocs.io/en/stable/error_code_list.html\#require-annotation-if-variable-type-is-unclear-var-annotated}{var-annotated}: 0.9\%- Style Convention \\
          \href{https://mypy.readthedocs.io/en/stable/error_code_list.html\#notify-about-an-annotation-in-an-unchecked-function-annotation-unchecked}{annotation-unchecked}: 0.8\% - Style Convention \\ 
          \href{https://mypy.readthedocs.io/en/stable/error_code_list.html\#check-types-in-assignment-statement-assignment}{assignment}: 0.8\% - Wrong Usage of Objects \\
          Total: 2.9\%
        \end{tabular}\\
    \hline
    Black
    & 753
      & \begin{tabular}[t]{@{}l@{}}
          \href{https://black.readthedocs.io/en/stable/usage_and_configuration/the_basics.html\#l-line-length}{line-length}: 81.5\% - Rule Specification \\
          \href{http://black.readthedocs.io/en/stable/usage_and_configuration/the_basics.html\#t-target-version}{target-version}: 46.4\% - Environment Setup\\
          \href{https://black.readthedocs.io/en/stable/usage_and_configuration/the_basics.html\#exclude}{exclude}: 30.4\% - Scope Control \\
          \href{https://black.readthedocs.io/en/stable/usage_and_configuration/the_basics.html\#include}{include}: 22.2\% - Scope Control \\
          \href{http://black.readthedocs.io/en/stable/usage_and_configuration/the_basics.html\#s-skip-string-normalization}{skip-string-normalization}: 10.1\% - Rule Specification \\
          Total: 95.4\%
        \end{tabular}
      & \multicolumn{1}{c|}{---} 
      & \multicolumn{1}{c|}{---} 
      \\
    \hline
    \multicolumn{5}{p{1.8\textwidth}}{ 
      Note: \#Proj is the number of projects using each linter; Total is proportion of projects that use at least one option from the top-5 most frequently used options, enabled or disabled linting rules across all projects; `--' denotes that the linter does not support the corresponding configuration option/rules. 
    }
  \end{tabular}
}
  
  \vspace{-6mm}
\end{table*}

\noindent \textbf{(2) Option Configuration Preference across Projects}: The \textit{Top-5 Used Option} column in Table~\ref{tab:option_freq} shows each linter's five most frequently configured options. 
The top-5 used options appear in over 75\% of projects for every linter, with Ruff, Flake8, and Black exceeding 90\%, Pylint exceeding 80\%, and Mypy at 78.5\%. 
The widespread adoption reflects that projects have a strong consensus on linter configuration options. 

For all five Python linters, 16 of the 25 top-5 options (64\%) fall into the \textit{rule specification} category. These mainly specify which linting rules check or not check, or finely configure linting rules. 
In particular, the maximum line length option appears in four of the five linters (all except Mypy). 
Next, five options(e.g., per-file-ignores, exclude and include) are in \textit{scope control} category. 
These options appear in the top-5 lists of Ruff, Flake8, and Black, indicating a common need to fine-tune which files or code regions are linted. 
Finally, three options belong to \textit{environment setup}. All three specify the target Python version, indicating that most projects default to a runtime environment and adjust this only when necessary.
And only a single \textit{result handling} option appears for Pylint, suggesting that its projects place additional emphasis on customizing output.
Overall, we observe this option configuration hierarchy: rule specification first, then scope control, followed by environment setup and result handling.  
For each linter, the project option configuration preference is as follows: 
    \begin{itemize}[leftmargin=10pt]
        \item For Ruff, the top-5 configured options span three categories: the \textit{rule specification} (the 1st$\sim$3rd options),  \textit{environment setup} (the 4th option), and \textit{scope control} (the 5th option), accounting for over 50\%, 47.6\%, and 36.4\%, respectively. 
        For the \textit{rule specification} category, most projects commonly set the maximum line length (`line-length'), followed by checking (`select') or not checking (`ignore') some linting rules. 
    The \textit{environment setup} option (`target-version') and \textit{scope control} option (`per-file-ignores') are to set the Python version and to not check files for specific linting rules. 

        \item For Pylint, the top-5 configured options fall into two categories: \textit{rule-specification} category (the 1st$\sim$3rd, and the 5th options), and \textit{result-handling} category (the 4th option), with each used in over 50\% of projects. 
        In the \textit{rule-specification} category, three options (e.g., `good-names') are used to finely configure code style, and one option (`disable') is used to specify which linting rules not to check. 
        In the \textit{result handling} category, the `reports' option is to specify whether to print a full report or only the linting messages. 
        
    \item For Flake8, the top-5 configured options span two categories: \textit{rule specification} category (the 1st, 2nd and 5th options), and the \textit{scope control} category (the 3rd and 4th options). 
    The first three options are used by over 50\% of projects, and the latter two are used by around 20\%. 
    Within the \textit{rule specification} category, two options like `max-complexity' are used to finely configure code style, and one option `ignore' is to specify which linting rules not to check. 
    In the \textit{scope control} category, both options (e.g., `exclude') specify which files to ignore. 
    
    
    \item For Mypy, the top-5 configured options fall into two categories: \textit{rule specification} category (the 1st$\sim$3rd and 5th options), and \textit{environment setup} category (the 4th option). 
     The first option is used by about 50\% of the projects, and the remaining four by 30\%$\sim$35\%. 
     For \textit{rule specification} category, four options like `ignore\_missing\_imports' are used to finely configure type checking behavior.  
     For \textit{environment setup} category, similar to Ruff and Black, `python\_version' option is to specify Python version used to parse. 
     
    
    \item For Black, the top-5 configured options fall into three categories: \textit{rule specification} (the 1st and 5th options), \textit{environment setup} (the 2nd option), and \textit{scope control} (the 3rd and 4th options).  
The 1st option is used by over 80\% of projects, the 2nd to 4th options by approximately 20\%$\sim$46\%, and the 5th option by around 10\%. 
Within the \textit{rule specification} category, all the two options are to finely configure code styles. 
For the \textit{environment setup} category, similar to Ruff and Mypy, the `target-version' is also to specify Python version used to parse. 
For \textit{scope control}, similar to Flake8, the two options (e.g., `exclude') are to include or exclude specific files or directories from linting.
    
    \end{itemize}

\vspace{-1mm}
\begin{tcolorbox}[colframe=black, width=\linewidth,boxrule=1pt,  boxsep=2pt,left=1pt,right=1pt,top=1pt,bottom=1pt, sharp corners]
\textbf{Finding 7:} Projects share common configuration priorities across the five Python linters. 
\textit{Rule specification} and \textit{scope control} dominate, reflecting common needs to customize checks and analysis scope.
In contrast, \textit{environment setup} and \textit{result handling} feature less prominently among commonly configured options, suggesting that their settings can generally remain at defaults unless needed for Python version compatibility or reporting.

 \end{tcolorbox}

\noindent \textbf{(3) Linting Rule Preference across Projects}: When projects select linting rules, they usually enable and disable a list of linting rules. 
Therefore, for each linter, we list the \textit{Top-5 Enabled Linting Rules} and the \textit{Top-5 Disabled Linting Rules} across projects in Table~\ref{tab:option_freq}.

\noindent \textbf{Enabled rule preference across projects}: For enabled linting rule, only projects using Ruff show a common preference: the top-5 enabled linting rules account for 75.2\% of projects. 
And projects prefer to enable core rule groups rather than individual linting rules. 
However, the top-five enabled linting rules for each of the other three linters appear in fewer than 16\% of projects, indicating that enabled rule selection is highly project-specific for the three linters (Pylint, Flake8 and Mypy). 

Black remains an outlier, as it does not support enabling linting rules. 

\noindent \textbf{Disabled rule preference across projects}: In contrast, disabling linting rules shows pronounced agreement for three linters: Ruff (48.3\%), Pylint (62.7\%), and Flake8 (62.9\%) frequently suppress strict style checks. 
In comparison, Mypy's top-five disabled rules appear in only 2.9\% of projects. 
Black still remains an outlier, as it does not support disabling rules. 

\noindent \textbf{Linting rule preference for each linter}: For each linter, the project configuration preferences for enabled rules and disabled rules are as follows: 
    \begin{itemize}[leftmargin=10pt]
        \item For Ruff, 
    rather than enabling specific linting rules, projects tend to enable core rule groups (e.g., `F', `E' and `I'), which accounts for 75.2\%.  
    The result indicates projects prefer broad rule coverage with minimal configuration overhead. 
    By contrast, when disabling rules, projects most often suppress strict linting rule checks in \textit{style convention} category like `E501' for line too long. 
    It indicates that projects prioritize flexibility over rigid adherence in \textit{style convention}. 

        \item For Pylint, projects do not show a clear preference for enabled rules: although all the top five enabled rules are individual rules, they account for only 10.5\% of projects, indicating the enabled rule of Pylint is diverse and project-specific. 
        In contrast, there is a clear preference for suppressing specific linting rules, with the top-5 disabled rules accounting for over 60\% of projects. 
    These disabled rules mainly focus on \textit{style convention} and  \textit{best practice}. 
    
    \item For Flake8, 
    enabled rules reveal no clear preference: although the five most enabled rules are rule groups, they appear in only 15.6\% of projects. 
    In contrast, projects consistently disable \textit{style convention} rule checks, which make up 62.9\% of projects.  
    
    \item For Mypy, projects do not exhibit a clear preference for enabled rules: the top-5 enabled rules are individual rules, yet they appear in only 5.0\% of projects, indicating a diverse range of configurations. 
    Similarly, there is no clear preference for disabled rules, where the top-5 disabled rules account for only 2.9\%. 
    This underscores the diverse, heterogeneous nature of Mypy configurations across projects. 

    \item For Black, 
    unlike the other four linters, Black does not support enabling or disabling individual linting rules, reflecting its opinionated design that prioritizes consistency and simplicity in code style. 
    \end{itemize}

\vspace{-3mm}
\begin{tcolorbox}[colframe=black, width=\linewidth,boxrule=1pt,  boxsep=2pt,left=1pt,right=1pt,top=1pt,bottom=1pt, sharp corners]
\textbf{Finding 8:} Among the five Python linters, only Black prohibits enabling or disabling rules, reflecting its opinionated design. 
Except for Black, when enabling rules, most linters (Pylint, Flake8 and Mypy) used by projects exhibit project-specific preferences. 
Conversely, when disabling rules, most linters (Ruff, Pylint and Flake8) used by projects show strong consensus, primarily suppressing \textit{style convention} checks.
This indicates that while most projects have unique linter needs, there is a consistent preference for less strict style conventions. 

 \end{tcolorbox}

%% file: discussion.tex



\subsection{Implications for Linter Users and Researchers} 
We further discuss valuable insights for the linter users and researchers based on the findings of our research questions. 

\noindent \textbf{Suggestions for linter users:} When selecting a linter from the five Python linters, users should base their choice on the project's specific needs. 
Ruff and Pylint are ideal for projects requiring comprehensive linting rules and high configurability, while Flake8 and Black are best for users seeking simplicity and minimal configuration overhead. 
Mypy, on the other hand, is particularly focused on type checking for object usage. 

When speed is critical, Ruff is the best choice, offering fast linting with running time typically under one second. 
Pylint requires more attention, as it can take over 1,000 seconds depending on the configuration. 
Users must balance between configuring linting rules and performance. 
For the other three linters (Flake8, Black and Mypy), running times generally remain within 10 seconds. 

For linter configuration, users should prioritize using \textit{pyproject.toml} to ensure consistency and compatibility across multiple linters. Among the available options, they can focus on configuring \textit{rule specification} and \textit{scope control}, which dominate the most frequently configured options across linters. \textit{Environment setup} and \textit{result handling} can generally remain at their default settings unless necessary.

\noindent \textbf{Future directions for researchers:} As shown in Section~\ref{linting_rule} and Section~\ref{config_prefer}, 44\% of linting rules are shared across multiple Python linters, and 56.6\% of projects use more than one linter. 
This highlights the potential for improving linter interoperability. 
As LLMs excel at understanding natural language, researchers can leverage LLMs to develop a unified framework that enables linters to work seamlessly together, reduces redundant checks, and enhances efficiency. 
Our dataset on rule overlap can serve as a benchmark. 
Besides, as new Python standards and tools (such as evolving versions of Python or emerging libraries) are introduced, it is crucial to examine how well linting rules cover these standards and how linters adapt to these changes, ensuring linters remain up-to-date and capable of enforcing the latest standards. 

For speed, some linters, especially Pylint, are costly in terms of running time on projects. 
Future work could leverage our comprehensive running time dataset to pinpoint performance bottlenecks and guide targeted optimizations for performance. 

Finally, projects using different linters adopt various configuration methods, and most projects have specific needs when selecting linting rules. 
Given the variety of linters, configuration methods and linting rules, automating configuration generation based on user requirements is essential to reduce their burden. 

\subsection{Threats to Validity}
One \textbf{internal threat} to validity is the potential personal bias in categorizing linting rules and configuration methods. 
To mitigate this, two annotators with over five years of linter usage experience independently analyzed the data and then discussed to reach a consensus. 
We release the code and data~\cite{data_code} for public review. 

An \textbf{external threat} lies in the scope of our study being limited to five Python linters. 
However, these linters are widely adopted, offer broad rule coverage, and provide distinct functionalities, making them representative. 
Our methodology offers a foundation for generalizing to other linters. 
In future work, we plan to extend our analysis to more Python linters and linters in other programming languages.

%% file: related.tex
Linters are static analysis tools that automatically detect potential code issues~\cite{relationship_linter_maintain,tomasdottir2020adoption,liter_use_warn}, helping developers maintain code quality by identifying errors early in the development process. 
These linters play a crucial role in ensuring code quality, consistency, and adherence to coding standards. 
Many popular programming languages have well-known linters, including Black~\cite{black} for Python, Checkstyle~\cite{ckstyle} for Java, and ESLint~\cite{eslint} for JavaScript. 
Current research on linters primarily falls into two categories. 

The one focuses on enhancing linters' ability to detect and fix code issues~\cite{latappy2023mlinter,eghbali2025dylin,hart2023eastwood,oumarou2015identifying,tsantalis2008jdeodorant,loriot2022styler,vayadande2023let,holden2024code,walchshofer2024introducing}. 
Latappy et al.~\cite{latappy2023mlinter} leveraged CodeBERT to learn coding practices, achieving good recall but low precision. 
Eghbali et al.~\cite{eghbali2025dylin} proposed DyLin, the first dynamic linter for Python, which detects 15 Python-specific antipatterns that are difficult to identify through static analysis. 
Holden et al.~\cite{holden2024code} applied a large language model to classify if code has issues or not. 
They found that model-based linter can achieve over 80\% accuracy on code issue classification. 

The other involves empirical studies on linters~\cite{beller2016analyzing,zampetti2017open,christakis2016developers,tomasdottir2020adoption,tomasdottir2017and,liter_use_warn,relationship_linter_maintain}.  
Tómasdóttir et al.~\cite{tomasdottir2020adoption} investigated developers' perceptions of ESLint by interviewing 15 developers and surveying 337 participants. 
They identified five key reasons why developers use ESLint and highlighted that a major challenge lies in selecting rules. 
Tjaša et al.~\cite{liter_use_warn} conducted a longitudinal case study of 15 JavaScript projects using linters. 
The results show that choices of linter and their configurations are rarely subjected to change once introduced to the project. 

Different from previous studies that primarily focus on JavaScript and a single linter such as ESLint, our work conducts a comprehensive empirical study on five popular Python linters. 
We systematically compare their rule checks, performance, configuration methods, and linter preferences to help developers better select and use linters.